\documentclass[11pt,twoside]{article}  
\usepackage{apn3conf}
\usepackage{lscape}

\def\nii{[N {\sc ii}]}
\def\sii{[S {\sc ii}]}

\def\oiii{[O {\sc iii}]}
\def\cliii{[Cl {\sc iii}]}
\def\ha{H$\alpha$}
\begin{document}   

\title{The Physical Parameters and Excitation of Jets and Knots in PNe}
\titlemark{Jets and Knots in PNe}

\author{D. R.\ Gon\c calves and A.\ Mampaso}
\affil{Instituto de Astrof\'{\i}sica de Canarias,
       \\E-38205 La Laguna, Tenerife, Spain, Email: denise@ll.iac.es}

\author{R.M.L.\ Corradi}
\affil{Isaac Newton Group of Telescopes,
       \\ Apartado de Correos 321, E-38700 La Palma, Spain}

\author{M.\ Perinotto}
\affil{Dipartimento di Astronomia e Scienza dello Spazio, 
       \\Universit\`a di Firenze,
       \\Largo E. Fermi 5, 50125 Firenze, Italy}

\contact{Denise R. Gon\c calves}
\email{denise@ll.iac.es}

\paindex{Gon\c calves, D. R.}
\aindex{Mampaso, A.}
\aindex{Corradi, R. M. L.}
\aindex{Perinotto, M.}

\authormark{Gon\c calves et al. }


\begin{abstract}          
We are carrying out a study of the physical parameters and excitation
of low-ionization structures (LIS) in planetary nebulae (PNe). Since
the optical morphology and kinematics of the LIS and the main
components (rims, shells and haloes) of our sample were studied
previously, our main goal now is to search for: i)~the density
contrasts between jets/knots and the main nebular components;
ii)~their main excitation processes; and iii)~their chemical
abundances. The first results of this survey --- based on the analysis
of NGC~7009, NGC~6543, NGC~6891 and K~4-47 --- are that there is no
significant density contrast between LIS and their surroundings, and
that most of the LIS studied (but not all) are mainly photoionized,
rather than shock excited.
\end{abstract}

\section{Introduction}

Our study is based on a sample of PNe with high quality images that
allow us to separate the properties of jets/knots from those of the
main components. Previous studies (Balick et al.~1994; Hajian et
al.~1997) have attempted to access the nature of FLIERs (fast,
low-ionization, emission regions) with similar analyses. We are now
trying to extend the work to a larger sample of LIS (20 PNe),
including not only FLIERs but also the other types of LIS. So far, we
have analyzed only part of the sample and we describe our results
here.

We obtained long-slit spectra of NGC~7009, NGC~6891, NGC~6543, and
K~4-47 on August 2001 using the 2.5 m Isaac Newton Telescope at La
Palma, Spain.  We used the IDS, with a configuration such that the
spectra cover the wavelength region from 3650 \AA\ to 7000 \AA, with
spectral and spatial sampling of 3.3 \AA~pix$^{-1}$ and
0.70$''$~pix$^{-1}$. The spectra were wavelength and flux calibrated,
corrected for extinction, and then emission line intensities were
measured allowing us to derive the physical parameters.

Note, from Table~1 of Gon\c calves et al.~(2001), that for most of the
PNe in our sample good kinematical studies exist in the
literature. They allow to compare the kinematical ages of the LIS with
those of the main nebular shells, and in particular of the bright
inner rims (created by the interaction of the AGB with the post-AGB
winds). For two of the PNe (NGC~7009 and NGC~6891), the LIS and rims
appear to be coeval; in NGC~6543 LIS are younger than the rim, and
the pair of knots in K~4-47 is likely older than the nebular core. Generally,
their orientations in space seems to coincide with the main axis of
the axisymmetrical shells (`polar' flows), or a little tilted (by as
much as 30$^{\circ}$ for those discussed here). With the exception of
NGC~6891, the tips of all the other jets are located (in projection)
outside the rim.

\section{NGC~7009 and the Other PNe}

\setcounter{footnote}{3}

In Table~1 we present the physical parameters ($N_{\rm e}$\sii,
$N_{\rm e}$\cliii, $T_{\rm e}$\oiii, $T_{\rm e}$\nii, and $T_{\rm
e}$\sii) for selected features of each PN. Note that ``R" stands for
rim, ``S" for the shell attached to the rim, ``J" for jet, ``K" for
knot, and ``NEB" for the results associated with the emission
integrated along the slit for the entire nebula. Position angles
(P.A.) are also given in Table~1 (numbers that appear beside the PN or
feature name).

\begin{figure}
\epsscale{1.0}
\plotone{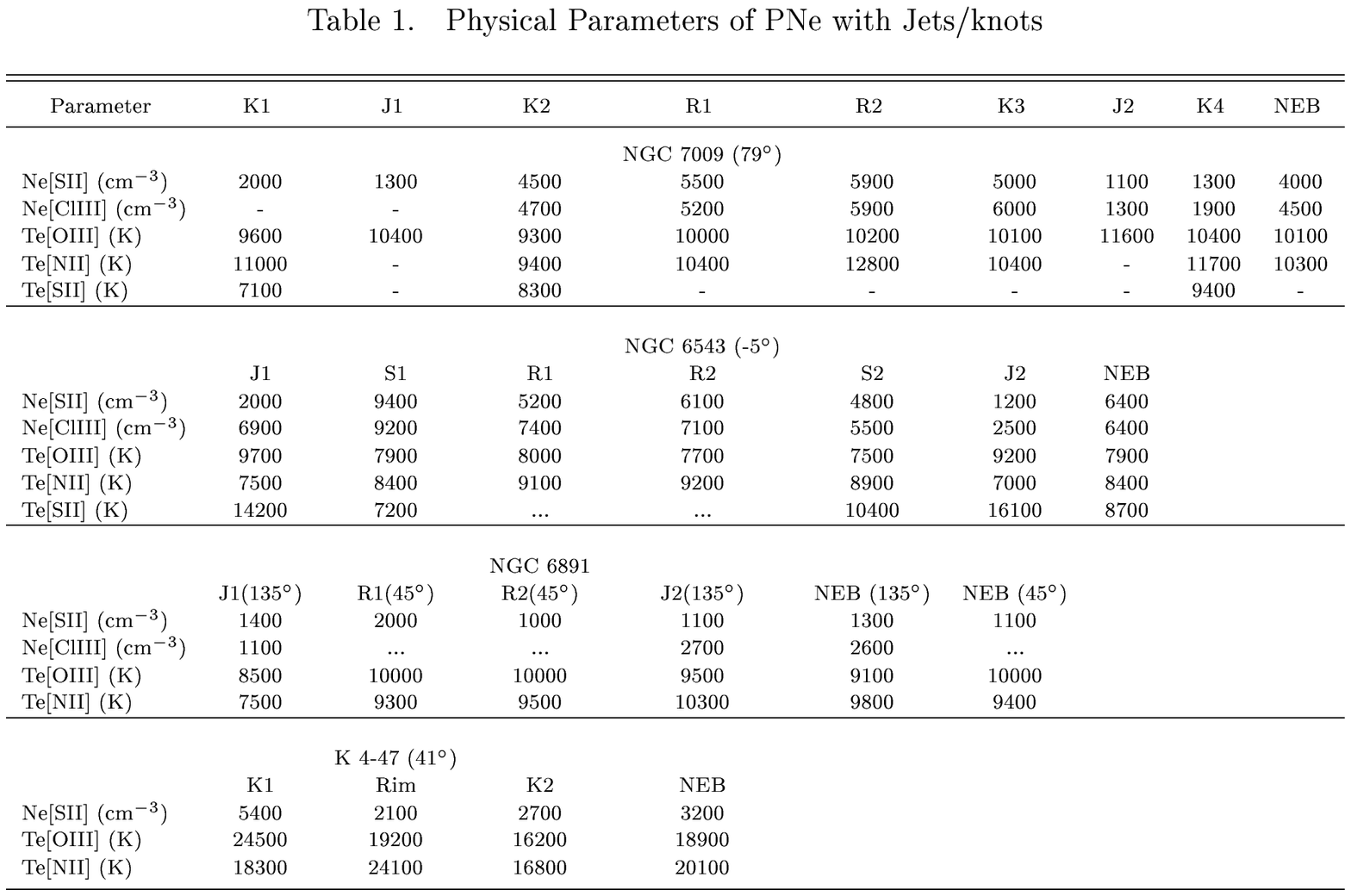}
\end{figure}

\setcounter{footnote}{0}

\smallskip\par\noindent {\it NGC~7009}: We selected six
`microstructures' in this PN (see Figure 2 of Gon\c calves's paper, this
volume)--- including the outer pair of knots (K1, K4), the pair of
jets (J1, J2), the inner pair of knots (K2, K3), in addition to the
rim (R1, R2).  Our main results are: i)~the
electron temperature throughout the nebula is remarkably constant,
$T_{\rm e}$\oiii = 10\,200~K; ii)~the bright inner rim and inner pair
of knots have similar densities of $N_{\rm e} \approx$ 6000~cm$^{-3}$,
whereas a much lower density of $\sim$1500~cm$^{-3}$ is derived for
the outer knots and for the jets; iii)~all the regions 
(rim, inner knots, jets and outer knots) 
are mainly radiatively excited; and
iv)~there are no clear abundance changes across the nebula for He, O,
Ne, or S.  There is a marginal evidence for an overabundance of
nitrogen in the outer knots (ansae), but the inner ones (caps) and the
rim have similar N/H values that are at variance with previous
results.

The derived values for $T_{\rm e}$, $N_{\rm e}$ and the abundances of
the whole nebula (see NEB column in Table 1) are in fair agreement
with previous determinations (Hyung \& Aller 1995; Rubin et al.~2002).
The properties of each region are more difficult to compare with other
works because there are no previous studies of this PN which make such
a detailed study of individual features of NGC~7009. But, again, for
the positions for which these parameters are available, there is at
least a marginal agreement with our values (Bohigas et al.~1994;
Balick et al.~1994).

The fact that none of NGC 7009's features is shock-excited put in
doubt the nature of its pair of jets.  Are these real jets that are
expanding supersonically through the halo?  
The failure to find evidence for shock excitation in the outer knots and
jets might simply reflect their moderate velocities or, as discussed
by Gon\c calves et al.~(2003), be in contradiction with the model
predictions.  However, very recently, two groups of researchers have
studied in detail the kinematics of this PN and both find that the
jets are indeed supersonic features (Fern\'andez et al.\ 2003;
F. Sabbadin, private communication). 

\smallskip\par\noindent {\it NGC~6543}: At variance with the other PNe
in the sample, the electron temperatures of its jet are higher than
those of the inner nebula, at least the [S II] temperature
\footnote{$T_e$\sii\ is obtained from the
$I$(6716\AA+6731\AA)/$I$(4069\AA+4076\AA) line ratio.}, whereas
densities are lower or equal to those of the inner regions.  None of
the six selected features of this PN is dominated by shock excitation.

\smallskip\par\noindent {\it NGC~6891}: Spectra were obtained at two
position angles, 135$^{\circ}$ (along the jets) and 45$^{\circ}$ (from
which we analyze the properties of the rim).  Table~1 shows that
densities and temperatures do not vary (within the errors) at the
positions of the rim and the jets.  As in the diagnostic diagrams
discussed below, the emission of the rim and the jets is dominated by
photoionization by the central star.

\smallskip\par\noindent {\it K~4-47}: This is a particularly unusual
PN, being mainly composed of an unresolved core and a pair of high
velocity blobs, with important differences when compared to the other
objects in the sample. First, note that its $T_{\rm e}$ are very much
higher than those of the others, $T_{\rm e}$\oiii\ always being above
16\,000~K throughout the PN and reaching values as high as 24\,000~K
for K1. These temperatures indicate, by themselves, that associated
emission should be contaminated or even dominated by shock
excitation. If so, the temperature calculation itself would be
doubtful.  Second, the densities of the knots in K~4-47 are higher
than in the core, in such a way that K1 has $N_{\rm
e}$\sii=5400~cm$^{-3}$, almost three times the density of the nebular
core, while K2 is only 30\% denser than the core. Third, the knots of
this PN are the only structures in the whole sample that clearly show
line ratios characteristics of shock excited emission (see next
section).

\section{Diagnostic Diagrams}

We put in these diagrams the line emission ratios \ha/[N
II](6548+6583) versus \ha/[S II](6716+6531) and [O
III](4959+5007)/\ha\ versus [S II](6716+6531)/\ha\ 
(from Phillips \& Cuesta 1999) for the different regions of a number of PNe
with knots and jets (Figure 3 of Gon\c calves's paper, this volume,
which contains the PNe discussed here, as well as some others whose
data come from the literature).  Note that the LIS in He 2-429 and He
1-1 (Guerrero et al.~1999) do not have peculiar velocities with
respect to the PNe main bodies in which they are embedded and are
useful for comparing their excitation properties with those of the
high-velocity features.

These diagrams show that only certain features of K~4-47, KjPn~8, and
M~2-48 show evidence for shock excitation. K 4-47 and M 2-48 share
many properties with young PNe, such as M~1-16 (Schwarz 1992), being
probably the youngest PN in the sample. So far, therefore, our search
for the excitation mechanisms of LIS (jets/knots) indicate that in
general high-velocity jets/knots in evolved PNe might not be shock
excited any longer. Another important thing shown by these diagrams 
is that properties of
the low-velocity features cannot be distinguished from those of the
high-velocity ones. 

Dopita (1997) and Miranda et al. (2000) have already called our
attention to the fact that jets in PNe contain some contribution from
shock excitation; however, the main excitation of these jets is
radiative.  In any case, very strong shocks, like those present in the
jets of young stellar objects, clearly do not exist in PNe.  Another
important issue to investigate is whether the spatial resolution
presently available is able to resolve the shocked regions from the
generally photoionized extended nebular environment (M. Perinotto,
private communication).

\end{document}